\documentclass[preprint,superscriptaddress,eqsecnum,nofootinbib]{revtex4}
\pdfoutput=1
\usepackage{amssymb,amsbsy,bm,amsmath,psfrag,dsfont}
\usepackage{hyperref}
\usepackage{graphicx}
\usepackage{epsfig}
\usepackage{slashed}

\begin{document}

\newcommand{\be}{\begin{equation}}
\newcommand{\ee}{\end{equation}}
\newcommand{\bea}{\begin{eqnarray}}
\newcommand{\eea}{\end{eqnarray}}
\newcommand{\no}{\noindent}

\title{Off-Shell Supersymmetry}
\author{Chiu Man Ho} \email{cmho@msu.edu}
\affiliation{Department of Physics and Astronomy, Michigan State University, East Lansing, MI 48824, USA}
\author{Nobuchika Okada} \email{okadan@ua.edu}
\affiliation{Department of Physics and Astronomy, University of Alabama, Tuscaloosa, AL 35487, USA}

\date{\today}

\begin{abstract}

Supersymmetry does not dictate the way we should quantize the fields in the supermultiplets, and so we have the freedom to quantize the
Standard Model (SM) particles and their superpartners differently. We propose a generalized quantization scheme under which
a particle can only appear off-shell, while its contributions to quantum corrections are exactly the same as those in the usual quantum field
theory. We apply this quantization scheme solely to the sparticles in the $R$-parity preserving Minimal Supersymmetric Standard Model (MSSM).
Thus sparticles can only appear off-shell. They could be light but would completely escape the direct detection at any experiments such
as the LHC. However, our theory still retains the same desirable features of the usual MSSM at the quantum level. For instance, the gauge
hierarchy problem is solved and the three MSSM gauge couplings are unified in the usual way. Although direct detection of sparticles is impossible, their existence can be revealed by precise measurements of some observables (such as the running QCD coupling) that may receive quantum corrections from them and have sizable deviations from the SM predictions. Also the experimental constraints from the indirect sparticle search are still applicable.

\end{abstract}

\maketitle

\section{Introduction}
Supersymmetric (SUSY) extension of the Standard Model (SM) is one of the most promising ways to solve the gauge hierarchy problem in the SM.
The Minimal Supersymmetric Standard Model (MSSM) not only offers a solution to the gauge hierarchy problem, but also provides
several interesting features for particle physics phenomenology. For example, the three MSSM gauge couplings are beautifully unified
at $M_{\rm GUT} \simeq  2 \times 10^{16}$ GeV, which suggests an interesting paradigm, Grand Unified Theory (GUT). With a conserved $R$-parity,
the lightest sparticle (neutralino) is a primary candidate for the dark matter in the Universe. Besides, electroweak symmetry breaking can be  triggered radiatively from a large top Yukawa coupling in the presence of SUSY-breaking terms. The SM-like Higgs boson mass is then predicted
as a function of soft SUSY-breaking terms for the third generation squarks.

It has been expected that SUSY is realized most naturally at the TeV scale, and hence sparticles can be discovered in the first run of the
Large Hadron Collider (LHC). In spite of a lot of efforts by the LHC, no signal of sparticles has been observed,
and the direct sparticle search has been postponed to the LHC Run II with a collider energy $13-14$ TeV. For a summary of the sparticle search
at the LHC, see \cite{SUSY_ATLAS, SUSY_CMS}. The null search results for colored sparticles such as the gluino $\tilde g$ and squarks $\tilde q$ imply lower bounds on their masses with roughly $m_{\tilde g} \gtrsim 1$ TeV and $ m_{\tilde q} \gtrsim $ 600 GeV. The lower bounds on non-colored sparticles such as sleptons, charginos and neutralinos are not so severe so far, with roughly a few hundred GeV.

In principle, it is possible that sparticles might have actually been produced at the LHC but have escaped from the detection because
their signals are too difficult to be distinguished from the SM background. This situation occurs when the energy of hadronic/leptonic jets
and the missing transverse energy associated with the sparticle cascade decay are significantly reduced in some SUSY models, such as the so-called Compressed SUSY~\cite{Compressed,Compressed2}, Stealth SUSY~\cite{Stealth,Stealth2} and $R$-parity violating SUSY models~\cite{RpV}.
For these models, the lower bounds on sparticle masses can be relaxed up to a few hundred GeV. However, it is not easy to construct such models in a natural way~\cite{SUSYRev1,SUSYRev2}. These models cannot completely hide the direct signal of sparticles forever. With the substantially
improved sensitivity at the LHC Run II, it is conceivable that they will be probed and constrained soon.

Contrary to the SUSY models which can tentatively hide the sparticles from direct detection, we would like to provide an $R$-parity preserving
MSSM scenario that can survive even if sparticles may never be directly observed at the LHC. Our guiding principle is to retain simplicity
and naturalness as much as possible. This work is also based on our observation that we do not need on-shell sparticles in order to
solve the gauge hierarchy problem and unify the three MSSM gauge couplings.

In this paper, we propose a generalized quantization scheme under which a particle can only appear off-shell. We then apply
this scheme to quantize the sparticles in the $R$-parity preserving MSSM while the SM particles are quantized in the conventional way.
Thus sparticles can only appear off-shell.\footnote{After this paper appeared on arXiv, we were informed that this
possibility was mentioned in \cite{Ghilencea}.} Without introducing any new interactions or exacerbating the naturalness, this evades all the direct detection bounds on sparticles. However, the contributions from the sparticles to quantum corrections in our theory are identical to those in the usual quantum field theory (QFT). Therefore, our MSSM retains the same desirable features in terms of quantum corrections as the usual MSSM. For instance, the gauge hierarchy problem is solved and the three MSSM gauge couplings are unified in the usual manner. The experimental constraints from the indirect sparticle search are still applicable and the same as in the usual MSSM. As a result, even though sparticles can only appear off-shell, SUSY is still broken with stringent bounds on flavor and additional CP violations \cite{Indirect}.

In the following sections, we first describe our generalized quantization scheme which leads to off-shell particles, and then we resolve
some apparent pathological issues associated with it. We apply this generalized quantization scheme solely to the sparticles in the
$R$-parity preserving MSSM, and so all the experimental bounds from the direct sparticle search disappear. Finally, we discuss the collider phenomenology of off-shell sparticles and possibilities to indirectly detect them.


\section{Generalized Quantization}
In the usual QFT, a real scalar field $\phi$ is quantized in the following way:
\bea
\phi(x) =\int\,\frac{d^3\,\mathbf{p}}{(2\pi)^{3/2}\,\sqrt{2\,\omega_\mathbf{p}}}\,\Big(\,a(\mathbf{p})\,e^{-ip x}+a^{\dagger}(\mathbf{p})\,e^{ipx}\,\Big)\,,~
\eea
where the annihilation operator $a(\mathbf{p})$ and the creation operator $a^{\dagger}(\mathbf{p})$ obey
the commutation relation $[\,a(\mathbf{p}),a^{\dagger}(\mathbf{p}')\,] =\delta^{(3)}(\mathbf{p}-\mathbf{p}')$.
This ensures that the equal-time canonical quantization scheme $[\,\phi(x),\,\dot{\phi}(x')\,] =i\,\delta^{(3)}(\mathbf{x}-\mathbf{x}')$
is satisfied. The vacuum $|\,0\,\rangle$ is chosen such that $a(\mathbf{p}) \,| \,0\, \rangle =0$.
As we know, if we compute the propagator for $\phi$, we will obtain the Feynman propagator.

Contrary to the conventional wisdom, we would like to introduce a generalized quantization scheme. We retain
the commutation relation $[\,a(\mathbf{p}),a^{\dagger}(\mathbf{p}')\,] =\delta^{(3)}(\mathbf{p}-\mathbf{p}')$, but
propose that
\bea
\label{lower}
a \,|\,n\,\rangle &=& \textrm{sign}\left(\,n-\frac12\,\right)\,\,\sqrt{\left|\,n-\frac12\,\right|}\,|\,n-1\,\rangle\,, \\
\label{raise}
a^\dagger \,|\,n\,\rangle &=& \sqrt{\left|\,n+\frac12\,\right|}\,|\,n+1\,\rangle \,,
\eea
for any integer $n$ which characterizes the number of energy units carried by the state $|\,n\,\rangle$.
Notice that for simplicity, we have suppressed the dependence on momentum $\mathbf{p}$ in
Eq. \eqref{lower} and Eq. \eqref{raise}.
(Apparently, this quantization scheme may lead to the disastrous negative energies and probably even worse problems.
We will hold on and resolve the apparent pathological issues below Eq. \eqref{half}. In fact, the resolution may
become self-evident when we move on.)
To elaborate, with the above choice, we have $a\, |\,0\,\rangle = -\frac{1}{\sqrt{2}}\,|-1\,\rangle$ and
$a^\dagger\, |\,0\,\rangle = \frac{1}{\sqrt{2}}\,|\,1\,\rangle$. This procedure implies that
\bea
\langle \,0\,| \,a^\dagger(\mathbf{p}) \,a(\mathbf{p}')\, | \,0 \,\rangle &=& -\frac12 \,\,\delta^{(3)}(\mathbf{p}-\mathbf{p}')\,, \\
\langle\,0\,| \,a(\mathbf{p}) \,a^\dagger(\mathbf{p}')\, |\,0\,\rangle &=& \frac12\,\,\delta^{(3)}(\mathbf{p}-\mathbf{p}')\,.
\eea
Thus, the vacuum expectation value of the Hamiltonian for $\phi$ without normal ordering, $H_\phi = 1/2\,\int\,d^3\,\mathbf{p}\, \,\omega_{\mathbf{p}} \,(\,a^\dagger(\mathbf{p}) \,a(\mathbf{p})+ a(\mathbf{p}) \,a^\dagger(\mathbf{p})\,)$, is
\bea
\langle\,0\,| \,H_\phi\, |\,0\,\rangle = 0\,.
\eea
In other words, the vacuum state $|\,0\,\rangle$ is no longer a state with the lowest energy but is simply a state with no particles and
thereby zero energy.

Applying the above quantization procedures, the propagator for $\phi$ turns out to be an average of the usual Feynman propagator with
an $+i \epsilon$ prescription and a similar one with an $-i \epsilon$ prescription:
\bea
\label{phi_propagator}
\langle 0| \,T\{\,\phi(x)\,\phi(y)\,\}\,| 0 \rangle
= \frac{1}{2}\,\Big(\, G_{+}(x-y) + G_{-}(x-y)  \,\Big)\,,~
\eea
where $G_{+}(x-y)$ and $G_{-}(x-y)$ are respectively given by
\bea
G_{+}(x-y) &=& \int\,\frac{d^4\,p}{(2\pi)^4}\,\, e^{-ip(x-y)}\, \frac{i}{p^2-m^2+i\,\epsilon}\,, ~~~\\
G_{-}(x-y) &=& \int\,\frac{d^4\,p}{(2\pi)^4}\,\, e^{-ip(x-y)}\, \frac{i}{p^2-m^2-i\,\epsilon}\,. ~~~
\eea

\begin{figure}[t!]
\includegraphics[height=2.5cm, width=6.0cm]{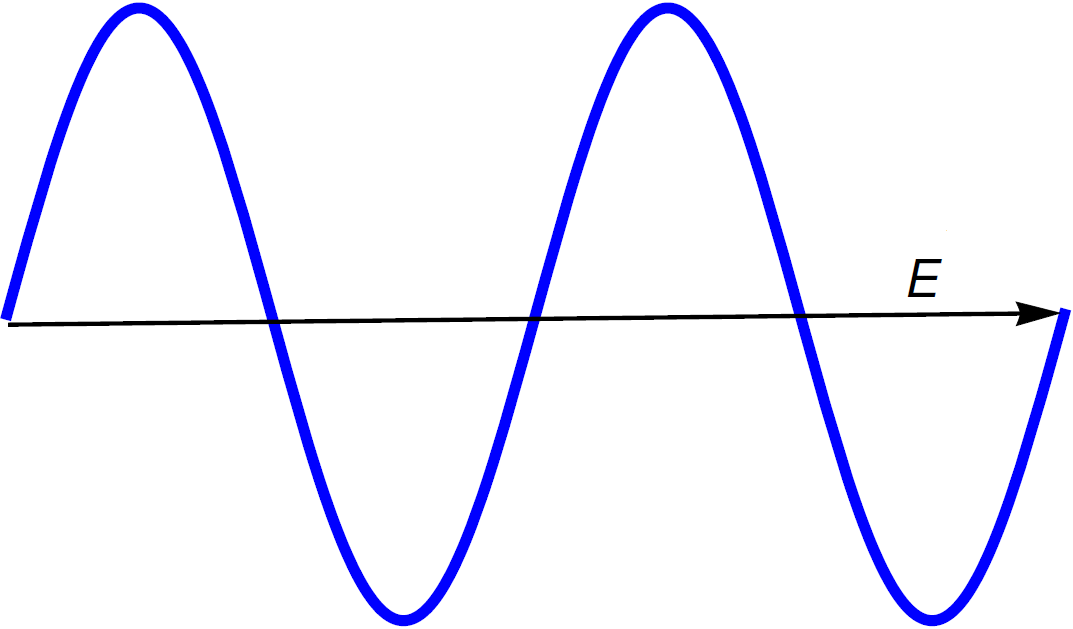}~~~~~~~~~~
\includegraphics[height=2.5cm, width=6.0cm]{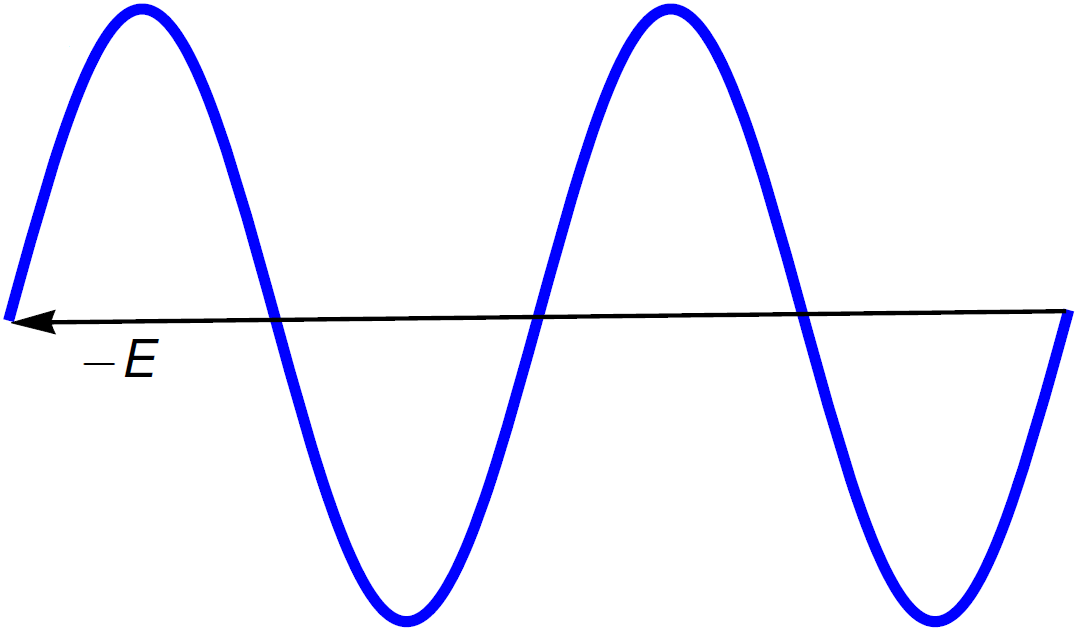} \\
\includegraphics[height=2.5cm, width=6.0cm]{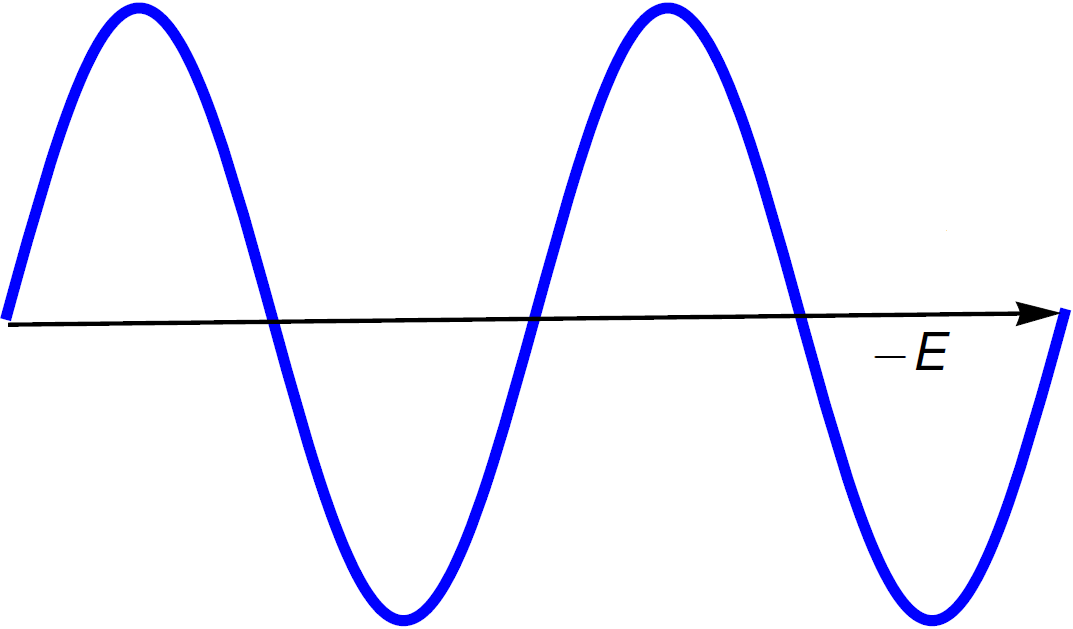}~~~~~~~~~~
\includegraphics[height=2.5cm, width=6.0cm]{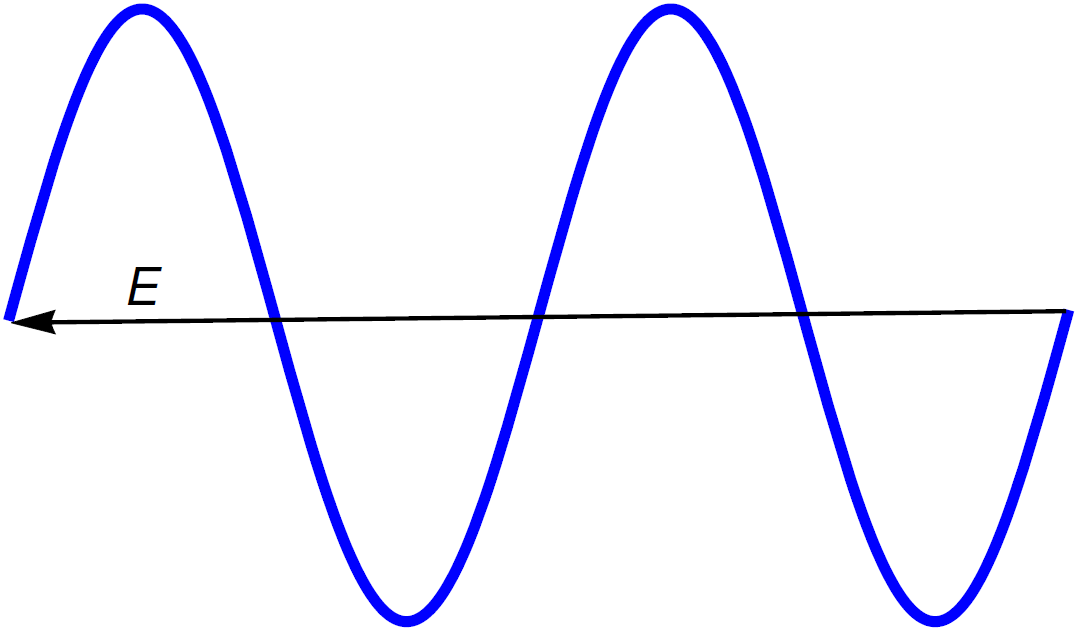}
\caption{Propagation of half-retarded and half-advanced energy modes, with the forward time direction pointing to the right. $+ E$ and $- E$ energy
modes are propagated both forward and backward in time with equal amplitudes.}
\label{fig:RetardedAdvanced}
\end{figure}

Actually, one can prove that the propagator, $\frac{1}{2}\left(\, G_{+}(x-y) + G_{-}(x-y)  \,\right)$, is equivalent to an average of the retarded and advanced Green's functions:
\bea
\label{half}
&& \frac{1}{p^2-m^2+i\,\epsilon} + \frac{1}{p^2-m^2-i\,\epsilon} \nonumber \\
&=& \frac{1}{2\,\omega_\mathbf{p} }\,\left(\, \frac{1}{p_0 - \omega_\mathbf{p} +i\,\epsilon} - \frac{1}{p_0 + \omega_\mathbf{p} -i\,\epsilon}     \,\right) + \frac{1}{2\,\omega_\mathbf{p} }\,\left(\, \frac{1}{p_0 - \omega_\mathbf{p} - i\,\epsilon} - \frac{1}{p_0 + \omega_\mathbf{p} +i\,\epsilon}     \,\right) \nonumber \\
&=& \frac{1}{2\,\omega_\mathbf{p} }\,\left(\, \frac{1}{p_0 - \omega_\mathbf{p} +i\,\epsilon} - \frac{1}{p_0 + \omega_\mathbf{p} +i\,\epsilon}     \,\right) + \frac{1}{2\,\omega_\mathbf{p} }\,\left(\, \frac{1}{p_0 - \omega_\mathbf{p} - i\,\epsilon} - \frac{1}{p_0 + \omega_\mathbf{p} -i\,\epsilon}     \,\right) \nonumber \\
&=& \frac{1}{(\,p_0+i\,\epsilon\,)^2-\omega_\mathbf{p}^2} + \frac{1}{(\,p_0-i\,\epsilon\,)^2-\omega_\mathbf{p}^2}\,.
\eea
Thus, it is also a half-retarded and half-advanced propagator. Under the generalized quantization scheme, the $\phi$ particle simultaneously propagates positive and negative energy modes both forward and backward in time with equal amplitudes. Therefore, the $\phi$ particle behaves like a ``standing wave" (in time) --- there is \emph{no} net energy flux being transferred. Since an on-shell particle carries a nonzero net energy flux, the $\phi$ particle propagated by the half-retarded and half-advanced propagator \emph{cannot} be on-shell. The corresponding physical picture is shown in Fig. \ref{fig:RetardedAdvanced}. In other words, the $\phi$ particle is ``ghost-like" in the sense that it can only appear off-shell. However, the procedure of making the $\phi$ particle ``ghost-like" is completely different from the usual Faddeev-Popov approach \cite{FP}. We do not need a ghost Lagrangian. What we need is simply the generalized quantization scheme described above.

Since the $\phi$ particle can only appear off-shell, the possibility of negative energies does not cause any problem. In fact, one can show
that if the state $|\,n\,\rangle$ has a negative energy with $n< 0$, it may also acquire a negative norm:
\bea
\label{NegativeNorm}
\langle\, n\,|\,n\,\rangle \, < 0\,, ~~~~~~ \textrm{if} \,\,\, n< 0 \,\,\,\textrm{and}\,\,\,|n|= \,\textrm{odd} \,,
\eea
which implies the existence of an indefinite metric.\footnote{The idea of indefinite metric was first invented by Dirac \cite{Dirac} and then elaborated by Pauli \cite{Pauli} in 1940s in an attempt to remove the divergences and construct a finite theory of quantum electrodynamics. Their attempt turned out to be not very satisfactory. But the canceling effect due to indefinite metric inspired Lee and Wick to construct an alternative finite theory of quantum electrodynamics \cite{LeeWick,LeeWick2}. The possible issues of Lee-Wick theory were discussed in \cite{CLOP, Nakanishi, Gross}. For a comprehensive lecture and review on indefinite metric QFT, one can consult \cite{Coleman_Indefinite} and \cite{Indefinite} respectively. In any case, due to the success of renormalization, these attempts for taming the divergences were largely forgotten. Recently, the ideas in \cite{LeeWick,LeeWick2} have been revived to construct a Lee-Wick Standard Model \cite{LWSM, LWSM2, LWSM3} which has many interesting properties (including the possibility to solve the gauge hierarchy problem).} But again, since the $\phi$ particle can only appear off-shell,
it does not contribute to the unitarity sum of the scattering amplitudes. Hence, unitarity is preserved despite the possible existence
of negative-norm states. This is analogous to the Faddeev-Popov ghosts which have negative norms but can only appear off-shell.

Since we have retained $[\,a(\mathbf{p}),a^{\dagger}(\mathbf{p}')\,] =\delta^{(3)}(\mathbf{p}-\mathbf{p}')$ in spite of the generalized
quantization scheme introduced in Eq. \eqref{lower} and Eq. \eqref{raise}, it follows that the equal-time commutator
$[\,\phi(x),\phi(y)\,]|_{x^0=y^0}$
should vanish as usual. This means that micro-causality is preserved under our framework. Moreover, for spacelike distances with $(x-y)^2 \equiv
- r^2 < 0$, there exists a reference frame where $x^0-y^0 =0$. It is well-known that
\bea
G_{+}(x-y) = \frac{m}{4\pi^2 r}\, K_1(m\,r)\,, ~~~~\textrm{for}\,\,\, (x-y)^2 < 0\,,
\eea
where $K_1(z)$ is the modified Bessel function of second kind. Thus outside the light-cone, the usual Feynman propagator predicts that the
propagation amplitude is exponentially small but nonzero (as $r \rightarrow \infty $). In contrast, one can show that
\bea
&& G_{-}(x-y) = - \theta(x^0-y^0)\int\,\frac{d^3\,\mathbf{p}}{(2\pi)^3}\,\, \frac{1}{2\,\omega_\mathbf{p}}\,\,e^{-ip(y-x)}
-\theta(y^0-x^0)\int\,\frac{d^3\,\mathbf{p}}{(2\pi)^3}\,\, \frac{1}{2\,\omega_\mathbf{p}}\,\,e^{-ip(x-y)}\,, \nonumber \\
\eea
and hence we have
\bea
G_{-}(x-y) = -\, G_{+}(x-y), ~~~~~~\textrm{for\, spacelike\, distances\, with}\,\,\, (x-y)^2 < 0\,.
\eea
Therefore, our propagator, which is half-retarded and half-advanced, is identically zero for spacelike distances.
In other words, the propagator obtained from our generalized quantization scheme predicts that the propagation amplitude is exactly
zero outside the light-cone and so it is truly causal.

In fact, our half-retarded and half-advanced propagator may have some resemblance to the absorber theory of radiation proposed by Feynman
and Wheeler \cite{Feynman, Feynman2}. They considered a half-retarded and half-advanced electromagnetic field in which electrons radiate symmetrically,
both forward and backward in time. As discussed in \cite{Coleman_Acausality}, due to subtle cancellations, the absorber theory is ``an apparently acausal theory that is not". Similarly, under our framework, each of the positive-energy and negative-energy states is propagated symmetrically, both forward and backward in time. The subtle ``destructive interference" renders the $\phi$ particle virtual (off-shell) and, at the same time, causal.

The extension of our generalized quantization scheme to a complex scalar field as well as a vector boson
is straightforward. (In the MSSM, we do not have spin-1 sparticles, so the extension to vector bosons is actually irrelevant.)
For a complex scalar given by
$\Phi(x)
=\int\,\frac{d^3\,\mathbf{p}}{(2\pi)^{3/2}\,\sqrt{2\,\omega_\mathbf{p}}}\,\left(\,a(\mathbf{p})\,e^{-ipx}+b^{\dagger}(\mathbf{p})\,e^{ipx}\,\right)$,
we require the usual commutation relations  $[\,a(\mathbf{p}),a^{\dagger}(\mathbf{p}')\,] = [\,b(\mathbf{p}),b^{\dagger}(\mathbf{p}')\,] =\delta^{(3)}(\mathbf{p}-\mathbf{p}')$ and impose that similar to $a(\mathbf{p})$ and $a^\dagger(\mathbf{p})$,\, $b(\mathbf{p})$ and $b^\dagger(\mathbf{p})$ acting on the state $|n\rangle$ satisfy the relations in Eq. \eqref{lower} and Eq. \eqref{raise} respectively. The vacuum expectation value of the Hamiltonian for $\Phi$ without normal ordering, $H_\Phi = 1/2\,\int\,d^3\,\mathbf{p}\, \,\omega_{\mathbf{p}} \,(\,a^\dagger(\mathbf{p}) \,a(\mathbf{p})+ a(\mathbf{p}) \,a^\dagger(\mathbf{p}) + b^\dagger(\mathbf{p}) \,b(\mathbf{p})+ b(\mathbf{p}) \,b^\dagger(\mathbf{p})\,)$, is $\langle\,0\,| \,H_\Phi\, |\,0\,\rangle = 0$. Besides, the propagator for $\Phi$, $\langle 0 | \,T\{\,\Phi^\dagger(x)\,\Phi(y)\,\}\,| 0 \rangle$, is of the same form as Eq. \eqref{phi_propagator}.

We extend our generalized quantization scheme to a Dirac fermion:
\bea
\psi(x) &=& \int\,\frac{d^3\,\mathbf{p}}{(2\pi)^{3/2}\,\sqrt{2\,\omega_\mathbf{p}}}\,\sum_{s=\pm}\, \Big(\,c(\mathbf{p},s)\,u(p,s)
\,e^{-ipx} + d^{\dagger}(\mathbf{p},s)\,v(p,s)\,e^{ipx}\,\Big)\,.
\eea
Here, the creation and annihilation operators obey the usual anticommutation relations
$\{c(\mathbf{p},s),\, c^{\dagger}(\mathbf{p}',s')\} = \{d(\mathbf{p},s),\, d^{\dagger}(\mathbf{p}',s')\} = \delta^{(3)}(\mathbf{p}-\mathbf{p}')\,\delta_{ss'}$. The spinors $u(p,s)$ and $v(p,s)$ satisfy the usual orthogonality conditions
as well as the completeness relations: $\sum_s u(p,s)\, \bar{u}(p,s) = \slashed{p}+m$\, and\, $\sum_s v(p,s)\, \bar{v}(p,s) = \slashed{p}-m$.
We introduce the following quantization steps:
\bea
\label{lower1}
c ~~ \textrm{or} ~~ d \,\,|\,n\,\rangle &=& \sqrt{\left|\,n-\frac12\,\right|}\,|\,n-1\,\rangle\,, ~\\
\label{raise1}
c^\dagger ~~ \textrm{or} ~~ d^\dagger \,\,|\,n\,\rangle &=& \sqrt{\left|\,n+\frac12\,\right|}\,|\,n+1\,\rangle \,,~
\eea
where for simplicity, we have suppressed the dependence on momentum $\mathbf{p}$ and spin index $s$ in
Eq. \eqref{lower1} and Eq. \eqref{raise1}. These imply that
\bea
 \langle \,0\,| \,c^\dagger(\mathbf{p},s) \,c(\mathbf{p}',s')\, | \,0 \,\rangle =
\langle\,0\,| \,c(\mathbf{p},s) \,c^\dagger(\mathbf{p}',s')\, |\,0\,\rangle
= \frac12\,\,\delta^{(3)}(\mathbf{p}-\mathbf{p}')\,\,\delta_{ss'} \,,\\
 \langle \,0\,| \,d^\dagger(\mathbf{p},s) \,d(\mathbf{p}',s')\, | \,0 \,\rangle =
\langle\,0\,| \,d(\mathbf{p},s) \,d^\dagger(\mathbf{p}',s')\, |\,0\,\rangle
= \frac12\,\,\delta^{(3)}(\mathbf{p}-\mathbf{p}') \,\,\delta_{ss'}\,.
\eea
Notice that due to the anticommutation relations, a difference from the bosonic case is that only $ n = -1, 0, 1$ are allowed. One can verify
that the norm for each of these states is positive-definite.\footnote{To some extend, $|-1\,\rangle$ for fermions may be interpreted as a hole state. In contrast, for bosons with $n< 0$ and $|n| =$ odd, $|\,n\,\rangle$ acquires a negative-norm. It is then unclear whether this interpretation is valid for bosons. Besides, one may wonder about the corresponding superpartners for the negative-energy states. We note that SUSY is realized at the Lagrangian level, namely the interactions between quantum fields. In the last paragraph of Section III, we argue that SUSY is fundamentally broken in the basis states underlying the generalized quantization scheme. Thus it is not mandatory to worry about the superpartners for the negative-energy states.} The vacuum expectation value of the Hamiltonian for $\psi$ without normal ordering,
$H_\psi = \sum_{s=\pm1/2}\,\int\,d^3\,\mathbf{p}\, \,\omega_{\mathbf{p}} \,(\,c^\dagger(\mathbf{p},s) \,c(\mathbf{p},s) - d(\mathbf{p},s) \,d^\dagger(\mathbf{p},s)\,)$, is
\bea
\langle\,0\,| \,H_\psi\, |\,0\,\rangle = 0\,.
\eea
Similar to the scalars, this means that the vacuum state $|\,0\,\rangle$ is no longer a state with the lowest energy but is simply a state
with no particles and thereby zero energy.

The propagator for $\psi$ is also half-retarded and half-advanced, where (with spinor indices suppressed) the corresponding
propagator $D_{-}(x-y)$ with an $-i \epsilon$ prescription is given by
\bea
D_{-}(x-y) &=& \int\,\frac{d^4\,p}{(2\pi)^4}\,\, e^{-ip(x-y)}\, \frac{i\,(\slashed{p}+m)}{p^2-m^2-i\,\epsilon} \\
\label{fermionD}
&=& \theta(x^0-y^0)\int\,\frac{d^3\,\mathbf{p}}{(2\pi)^3}\,\, \frac{\slashed{p}-m}{2\,\omega_\mathbf{p}}\,\,e^{-ip(y-x)}
-\theta(y^0-x^0)\int\,\frac{d^3\,\mathbf{p}}{(2\pi)^3}\,\, \frac{\slashed{p}+m}{2\,\omega_\mathbf{p}}\,\,e^{-ip(x-y)}\,.\nonumber \\
\eea
The exact form of the propagator for $\psi$ is
\bea
&& \langle 0| \,T\{\,\psi_\alpha(x)\,\bar{\psi}_\beta(y)\,\}\,|0\rangle \equiv  S(x-y)_{\alpha\beta} \nonumber \\
&=& \int\,\frac{d^4 p}{(2\pi)^4}\, e^{-ip(x-y)} \,\left(\,\slashed{p}+m\,\right)_{\alpha\beta}\,\, \frac{1}{2}\,\left(\,\frac{i}{p^2-m^2+i\,\epsilon} + \frac{i}{p^2-m^2-i\,\epsilon} \,\right)\,,
\eea
and so the $\psi$ particle can only appear off-shell. Following the similar arguments for the scalars, the possibility of negative energies
is not a problem and micro-causality is preserved. Besides, one can verify, using Eq. \eqref{fermionD}, that this propagator is exactly zero outside the light-cone.

For a Majorana fermion with $\psi = \psi^{c} \equiv C\,\bar{\psi}^{T}$ where $C$ is the charge conjugation matrix, the quantization steps are similar except that we set $c(\mathbf{p},s) = d(\mathbf{p},s)$. There are two additional propagators:
\bea
\langle 0| \,T\{\,\psi_\alpha(x)\,\psi_\beta(y)\,\}\,|0\rangle &=& \left[\,C^{-1}\,S(x-y)\,\right]_{\alpha\beta}\\
\langle 0| \,T\{\,\bar{\psi}_\alpha(x)\,\bar{\psi}_\beta(y)\,\}\,|0\rangle &=& \left[\,-S(x-y)\,C\,\right]_{\alpha\beta}\,.
\eea

As we know, it is the integration contour that determines the amount of quantum corrections due to the propagators. In the usual QFT, the
propagators behave like $\frac{1}{p^2-m^2+i\,\epsilon}$ whose poles are in the II \& IV quadrants. One can close the
contour of integration either in the upper or lower half-plane. With the generalized quantization, the propagators for particles behave like an
average of $\frac{1}{p^2-m^2+i\,\epsilon}$ and $\frac{1}{p^2-m^2-i\,\epsilon}$. Note that the propagator $\frac{1}{p^2-m^2-i\,\epsilon}$ has
poles in the I \& III quadrants. Thus, a particle with generalized quantization, \emph{as a single particle}, has poles in all of the four quadrants.

\begin{figure}[t!]
\includegraphics[height=6.5cm, width=6.5cm]{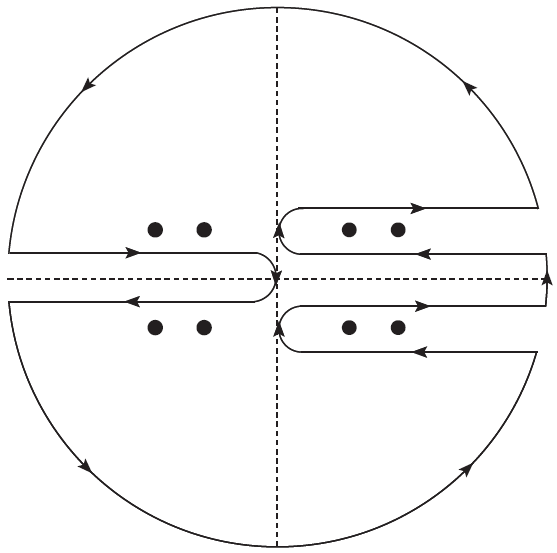}~~~~~~~~
\includegraphics[height=6.5cm, width=6.5cm]{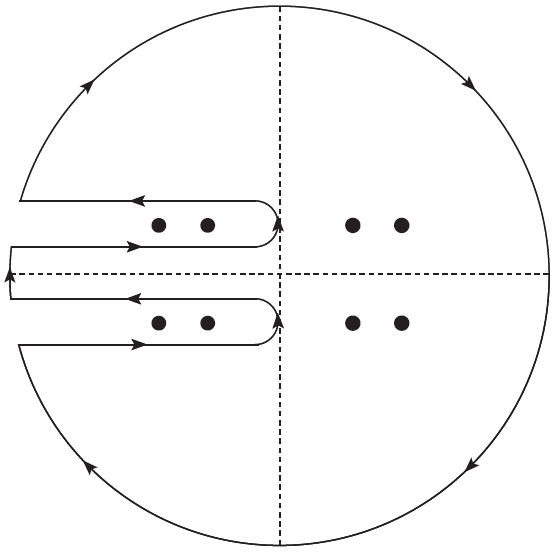}
\caption{The integration contours prescribed for particles with generalized quantization. The $x$-axis and $y$-axis represent $\textrm{Re}\,p_0$
and $\textrm{Im}\,p_0$ respectively. The symbols $\bullet$ denote
the possible poles (not drawn to scale). For the left contour, only the possible poles in the III \& IV quadrants are picked up.
For the right contour, only the possible poles in the I \& II quadrants are picked up.}
\label{fig:contour}
\end{figure}

Since particles with generalized quantization have unusual pole structures, it is conceivable that unconventional integration contours are
needed for them. We prescribe the integration contours for particles with generalized quantization in Fig. \ref{fig:contour}, where
the $x$-axis and $y$-axis represent $\textrm{Re}\,p_0$ and $\textrm{Im}\,p_0$ respectively. For the left contour, only the possible poles in
the III \& IV quadrants are picked up. For the right contour, only the possible poles in the I \& II quadrants are picked up.
Either of the left or right contour allows the poles of $\frac{1}{p^2-m^2+i\,\epsilon}$ and $\frac{1}{p^2-m^2-i\,\epsilon}$ to contribute constructively. With either contour in Fig. \ref{fig:contour}, one could verify that quantum corrections from particles with generalized
quantization are exactly the same as those in the usual QFT.

\section{SUSY with Generalized Quantization}
For a given supermultiplet, supersymmetry exhibited at the Lagrangian does not dictate the way we should quantize the fields. Providing the usual $R$-parity preserving MSSM Lagrangian, we propose that the SM particles are quantized in the conventional way, while their superpartners are quantized according to the generalized scheme described above.\footnote{It would be appealing if one could find a theoretical motivation for making this choice of quantization. A more fundamental theory would probably exhibit the property that the Hilbert spaces for the SM particles and their superpartners are different. However, this is beyond the scope of the current paper which concerns more about the phenomenology. We are actively working on this issue and hope to report elsewhere soon.} Therefore, sparticles can only appear off-shell in our theory, which evades the direct detection at any experiments. However, since quantum corrections from the sparticles are identical to those in the usual QFT, the gauge hierarchy problem is solved in the usual way.

With the soft SUSY-breaking terms, our MSSM scenario with sparticles obeying the generalized quantization leads to the same attractive phenomenological consequences as the usual MSSM does~\cite{SUSYrev}. For example, the three MSSM gauge couplings are unified at $M_{\rm GUT} \simeq 2 \times 10^{16}$ GeV, the electroweak symmetry is radiatively broken, and the SM-like Higgs boson mass is obtained through stop loop corrections. Although all the experimental constraints from the direct detection disappear, the results from the indirect sparticle search are still
applicable to our MSSM scenario to constrain the sparticle mass spectrum. Similar to the usual MSSM, this requires SUSY to be
broken with stringent bounds on flavor and additional CP violations.

As a consequence of applying the generalized quantization scheme to sparticles, we obtain $\langle \, 0\, |\,H_{\textrm{sparticles}}\,|\,0\,\rangle = 0$. This implies that $\langle\,0\,|\,H_{\textrm{SM}}+H_{\textrm{sparticles}}\,|\,0\,\rangle \neq 0$ even if SUSY is manifest.
Thus, it appears that SUSY is fundamentally broken in the basis underlying the generalized quantization scheme. However, the off-shell MSSM Lagrangian is exactly the same as the original MSSM Lagrangian including the soft terms. The structure of quantum corrections (in a typical SUSY theory) responsible for the potential cancellations of divergences is still retained, so the SUSY breaking in the vacuum state has no practical effect on particle physics phenomenology. We can simply remove the constant vacuum energy by normal ordering,
as in the usual QFT.

\section{Phenomenology of Off-Shell SUSY}
As mentioned above, although it is impossible to directly detect sparticles which can only appear off-shell, their existence can be
indirectly identified through their contributions to quantum corrections for some observables. For example, the fine structure constant at the Z-pole ($m_Z$) is very precisely measured as $ \alpha_{\rm em}(m_Z)^{-1}  = 127.918 \pm  0.019$~\cite{alpha}, which is consistent with the SM prediction
for the evolution of the fine structure constant from low energy to the Z-pole. If charged sparticles such as squarks, sleptons and charginos are involved in the evolution, the resultant fine structure constant at the Z-pole will be altered from the SM prediction.
This sets the lower bound on the charged sparticle masses as $\tilde{m} \gtrsim m_Z$.

The discussion about the fine structure constant may give us an idea about how to identify the existence of light colored (off-shell) sparticles such as the gluino and squarks at the LHC, even though the experimental data would show no indication of sparticle productions. At energies higher than their masses, the colored sparticles are involved in the running QCD coupling and will deflect the running from the trajectory predicted by the SM.
Employing the 1-loop renormalization group equation for the QCD coupling, we define a deviation of the running QCD coupling at an energy
scale $\mu$ as
\begin{eqnarray}
 \Delta (\mu) &\equiv& \frac{\alpha_{\rm s}^{\rm MSSM}(\mu)}{\alpha_{\rm s}^{\rm SM}(\mu)} -1   \\
 & \approx &
  \frac{\alpha_{\rm s}^{\rm SM}({\tilde m})}{2 \pi} (b_{\rm MSSM}-b_{\rm SM} ) \; \ln \left( \frac{\mu}{\tilde m} \right)\,,
\end{eqnarray}
where $\alpha_{\rm s}^{\rm SM}$ is the running SM QCD coupling, $\alpha_{\rm s}^{\rm MSSM}$ is the running QCD coupling with the
contributions from colored sparticles with a degenerate mass ${\tilde m}$,
and $b_{\rm SM}=-7$ ($b_{\rm MSSM}$) is the QCD beta function coefficient in the SM (MSSM).
Using $\alpha_{\rm s}(M_t)=0.0928$ with a top quark pole mass $M_t=173.34$ GeV~\cite{RGE},
we find $\Delta(1 \, {\rm TeV})\approx 3.6 \%$, for ${\tilde m}=500$ GeV
and $b_{\rm MSSM}=-3$ due to the contributions from the degenerate gluino and three generations of squarks.
This deviation is slightly smaller than the error of the current measurement of the QCD coupling
constant in the TeV range~\cite{QCDcoupling, QCDcoupling2}. However, we obtain $\Delta(\mu \gtrsim 3\, {\rm TeV}) \gtrsim 9 \%$ for
the same parameters. Therefore, a more precise measurement of the QCD coupling at sufficiently
higher energy and luminosity may reveal the off-shell colored sparticles.

\section{Conclusions and Discussions}
In this paper, we propose a novel MSSM scenario where sparticles are quantized under a generalized scheme to only appear off-shell. However, quantum corrections from the sparticles in this theory are exactly the same as those in the usual QFT.
As a consequence, most of the phenomenologically attractive properties of the usual MSSM, such as the solution to the gauge hierarchy problem and the successful gauge coupling unification, remain the same. Although direct detection of sparticles is impossible, their existence can be revealed through precise measurements of observables to which off-shell sparticles give sizable quantum corrections.

In principle, one could apply our generalized quantization scheme to any other theories so as to evade the corresponding bounds from
direct detection. Nevertheless, sparticles are particularly well-suited for the generalized quantization. The reason is that the most
important merit of sparticles is due to their off-shell quantum contributions. For instance, we do not need on-shell sparticles in order to solve the gauge hierarchy problem and unify the three MSSM gauge couplings.

Of course, the lightest sparticle (e.g. neutralino) would no longer be a viable dark matter candidate if it can only appear off-shell. However,
this should not be considered as a serious deficiency of our MSSM scenario. Providing a viable dark matter candidate is a just bonus of the
usual MSSM, and it is easy enough to construct other models that give a promising dark matter candidate. Superpsymmetry is most crucial for
solving the gauge hierarchy problem and unification of the three MSSM gauge couplings.


Our encouraging message is that even if none of sparticles is observed at the LHC Run II, there is still a hope that the sparticles are light but can only appear off-shell. In that case, their existence may have to be indirectly identified through a precise measurement of the running QCD coupling.

In fact, our work is more general than supersymmetry. Our idea has provided a new vision that some new physics may only appear off-shell. An intriguing collider signature of this kind of new physics is that we may see sizable deviations from the SM predictions at precision
measurements despite the absence of new particles at direct detection.

Finally, in the present work, the realization of the idea of off-shell supersymmetry relies on the generalized quantization scheme. In order to better justify the consistency of this idea, we have demonstrated that it could be formulated from the path-integral approach
in \cite{pathintegral}. \\

\vspace{3cm}

\emph{Acknowledgments.}\,\,
CMH thanks Arthur Kosowsky who first brought to his attention the papers by Feynman and Wheeler some years ago, which seeded
the inspiration to this work. CMH was supported by the Office of the Vice-President for Research and Graduate
Studies at Michigan State University.


\end{document}